\documentclass[aps,prd,showpacs,twocolumn,groupedaddress]{revtex4}

\usepackage{graphicx}

\begin{document}


\title{On the Direct Search for Spin-dependent WIMP Interactions}


\author{TA Girard}
\email[]{criodets@cii.fc.ul.pt}
\author{F. Giuliani}
\affiliation{Centro de F\'isica Nuclear, Universidade de Lisboa,
1649-003 Lisboa, Portugal}


\date{\today}

\begin{abstract}
We examine the current directions in the search for spin-dependent
dark matter. We discover that, with few exceptions, the search
activity is concentrated towards constraints on the WIMP-neutron
spin coupling, with significantly less impact in the WIMP-proton
sector. We review the situation of those experiments with
WIMP-proton spin sensitivity, toward identifying those capable of
reestablishing the balance.
\end{abstract}




\maketitle

\section{INTRODUCTION}

The direct search for weakly interacting massive particle (WIMP)
dark matter continues among the forefront efforts of experimental
physics. The search is largely motivated by the continuing absence
of a second positive signal with annual modulation confirming the
result of the DAMA/$NaI$ experiment despite significantly improved
detectors, and especially following the several recent reports of
possible indirect observation of dark matter annihilation
\cite{egret,hess,heat}.

Direct search efforts, based on the detection of nuclear recoils in
WIMP-nucleon interactions, have been traditionally classified as to
whether sensitive for the spin-independent or spin-dependent WIMP
channel \cite{Lewin}. While the former are generally constructed
simply on the basis of heavy target nuclei (since the interaction
cross section varies as $A^{2}$), the latter require consideration
of the spin structure of the detector nuclei and are customarily
defined by whether the primary experiment sensitivity is to
WIMP-proton or WIMP-neutron spin coupling. The main efforts to date
have been in spin-independent searches, generally because the
anticipated cross sections are larger owing to a coherent
interaction across the nucleus. As recently noted by Bednyakov and
$\check{S}$imkovic \cite{zero}, however, the importance of the
spin-dependent sector cannot however be ignored: such searches
provide twice stronger constraints on SUSY parameter space, permit
the detection of large nuclei recoil energy due to nuclear structure
effects in the case of heavy target nuclei, and prevent missing a
dark matter signal which might be suppressed in the spin-independent
sector. If the neutralino is predominantly gaugino or higgsino
states, the coupling is only spin-dependent \cite{gaitskell}.

The above distinction in search efforts has become somewhat blurred,
since many detector heavy isotopes also possess spin and even a
small natural isotopic abundance can produce significant
constraints. In fact, a single detector can simultaneously provide
restrictions on both channels of the WIMP-nucleon interaction, but
with sensitivity in each channel dependent on the nature of the
detector material. The most stringent limits in the spin-dependent
sector are currently provided by experiments traditionally
considered spin-independent.

The future thrust of direct searches for WIMP dark matter is defined
by a number of project upgrades and several new high profile
activities. Basically designed for deeply probing the
spin-independent phase space, these experiments project eventual
sensitivities beginning well below the controversial DAMA/$NaI$
result and extending to cross sections as small as $10^{-10}$ pb in
the WIMP-nucleon interaction. We here consider the impact of this
activity thrust on the spin-dependent sector, finding that of those
experiments with spin-sensitivity, most all will provide
increasingly improved restrictions predominantly on the possible
WIMP-neutron spin coupling, with significantly less impact on the
WIMP-proton coupling. This latter sector is in fact observed to have
previously received comparatively little direct experiment
attention, with the current restrictions derived from $NaI$
experiments already surpassed by two orders of magnitude in
sensitivity by the indirect searches \cite{superK,baksan}. Given
that there however remain important theoretical questions regarding
the extraction of the indirect results, we examine the situation of
direct experiments with predominantly WIMP-proton spin sensitivity,
towards identifying those with capacity to provide similar
restrictions.

Sec. II reviews the current experimental situation and thrust of new
initiatives in the search effort. The impact of these is discussed
in Sec. III, and summarized in Sec. IV.

\section{EXPERIMENTAL SITUATION}

\begin{figure}[h]
   \includegraphics[width=8 cm]{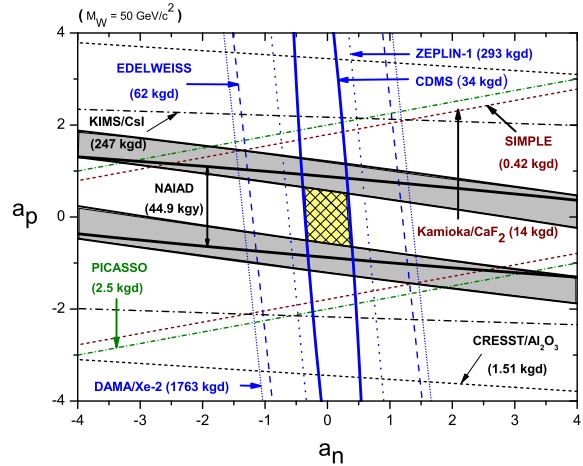}\\
   \caption{spin-dependent exclusions for various direct search activities, the region
   permitted by each experiment lying inside the respective band: DAMA/$Xe$-2 (small dot),
   EDELWEISS/$Ge$ (long dash), ZEPLIN-1/$Xe$ (big dot), CDMS/$Ge$ (solid), NAIAD/$NaI$ (solid),
   PICASSO/$C_{4}F_{10}$ (dash-dot), SIMPLE/$C_{2}ClF_{5}$ (short dash), Kamiokana/$CaF_{2}$ (short dash),
   CRESST-1/$Al_{2}O_{3}$ (short dash), KIMS/$CsI$ (dash-dot); the controversial positive result of
   DAMA/$NaI$ is shown as shaded. The unexcluded region, defined by the
   intersection of CDMS \cite{cdms06SD} and NAIAD \cite{naiad2}, is shown as
   crosshatched.}\label{SDexc}
\end{figure}

The situation for spin-dependent (SD) activity is shown in Fig.
\ref{SDexc} at $90\%$ C.L. for a WIMP mass (M$_{W}) = 50$
GeV/c$^{2}$, obtained in a model-independent, zero momentum transfer
approximation \cite{FGprl,Tovey}. The $a_{p},a_{n}$ are the
WIMP-proton (neutron) coupling strengths in the spin-dependent
WIMP-nucleus cross section

\begin{equation}\label{cross2}
{\sigma_{A}^{(SD)} = \frac{32}{\pi}G_{F}^{2}\mu_{A}^{2}
(a_{p}\langle S_{p}\rangle +a_{n}\langle S_{n}\rangle
)^{2}\frac{J+1}{J}} ,
\end{equation}

\noindent where $\mu_{p,n,A}$ =
$\frac{M_{W}m_{p,n,A}}{M_{W}+m_{p,n,A}}$ is the WIMP-proton
(-neutron,-nucleus) reduced mass, $\langle S_{p,n} \rangle $ is the
expectation value of the proton (neutron) group spin, $G_{F}$ is the
Fermi coupling constant,  and J is the total nuclear spin. The
figure is constructed from the published results of the respective
experiments, using

\begin{equation}\label{elli}
{\sum_{A}\left( \frac{a_{p}}{\sqrt{\sigma_{p}^{lim(A)}}}\pm
\frac{a_{n}}{\sqrt{\sigma_{n}^{lim(A)}}}\right)^{2}  \leq
\frac{\pi}{24G_{F}^{2}\mu _{p}^{2}} } ,
\end{equation}

\noindent with $\sigma_{p,n}^{lim(A)}$ the proton and neutron cross
section limits defined by

\begin{equation}\label{cross} {\sigma_{p,n}^{lim(A)} =
\frac{3}{4}\frac{J}{J+1}\frac{\mu_{p,n}^{2}}{\mu_{A}^{2}}\frac{\sigma_{A}^{lim}}{\langle
S_{p,n}\rangle^{2}} }  ,
\end{equation}

\noindent where $ \sigma_{A}^{lim}$ is the upper limit on $ \sigma
_{A}^{(SD)}$ obtained from experimental data, and the small
difference between $m_{p}$ and $m_{n}$ is neglected. The sum in Eq.
(\ref{elli}) is over each of the detector nuclear species, with the
sign of the addition in parenthesis being that of $\langle
S_{p}\rangle/\langle S_{n}\rangle$, and is an ellipse except in the
case of single-nuclei experiments for which the ellipse degenerates
into a band. When $\sigma_{p,n}^{lim(A)}$ were not available, they
have been obtained from published cross section limits as described
in Ref. \cite{FGprl,FGprd}.

At this magnification, with the exception of CDMS/$Ge$, the
exclusion plot is seen to consist of essentially horizontal
($a_{p}$-sensitive), vertical ($a_{n}$-sensitive) and diagonal
bands, within which lies the allowed area, the exterior being
excluded.

\begin{table*}
\caption{\label{tabisos}spin-sensitive detector isotopes and their
experiments.}
\begin{ruledtabular}
\begin{tabular}{ccccc}
  isotope & Z & J$^{\pi}$ & abundance (\%) & experiment\\ \hline
  $^{3}$He & 2 & 1/2$^{+}$ & $<<$1 & MIMAC \cite{mimac}\\
  $^{7}$Li & 3 & 3/2$^{-}$ & 93 & Kamioka/$LiF$ \cite{lif}\\
  $^{13}$C & 6 & 1/2$^{-}$ & 1.1 & PICASSO \cite{newpicasso}, SIMPLE \cite{plb2}, COUPP \cite{coupp}\\
  $^{17}$O & 8 & 5/2$^{+}$ & $<<$1 & ROSEBUD \cite{rosebudII}, CRESST \cite{cresst}\\
  $^{19}$F & 9 & 1/2$^{+}$ & 100 & SIMPLE \cite{plb2}, PICASSO \cite{newpicasso}, Kamioka \cite{lif,naf}, COUPP \cite{coupp}\\
  $^{21}$Ne & 10 & 3/2$^{+}$ & $<<$1 & CLEAN \cite{clean}\\
  $^{23}$Na & 11 & 3/2$^{+}$ & 100 & DAMA \cite{danai}, NAIAD \cite{naiad2}, ANAIS \cite{anais}, LIBRA \cite{libra}\\& & & & Kamioka/$NaF$ \cite{naf}\\
  $^{27}$Al & 13 & 5/2$^{+}$ & 100 & ROSEBUD \cite{rosebudII}\\
  $^{29}$Si & 14 & 7/2$^{+}$ & 4.7 & CDMS \cite{cdms06SD}\\
  $^{35}$Cl & 17 & 3/2$^{+}$ & 76 & SIMPLE \cite{plb2}\\
  $^{37}$Cl & 17 & 3/2$^{+}$ & 24 & SIMPLE \cite{plb2}\\
  $^{43}$Ca & 20 & 7/2$^{-}$ & $<<$1 & CRESST-II \cite{papcresstII}, Kamioka/$CaF_{2}$ \cite{CaF}\\
  $^{67}$Zn & 30 & 5/2$^{-}$ & 4.1 & CRESST-II \cite{papcresstII}\\
  $^{73}$Ge & 32 & 9/2$^{+}$ & 7.8 & HDMS \cite{hdms}, CDMS \cite{cdms06SD}, GENIUS \cite{genius2}, EDELWEISS \cite{eureka}\\
  $^{127}$I & 53 & 5/2$^{+}$ & 100 & DAMA \cite{danai}, NAIAD \cite{naiad2}, KIMS \cite{kims}, ANAIS \cite{anais}\\& & & & LIBRA \cite{libra}, COUPP \cite{coupp}\\
  $^{129}$Xe & 54 & 1/2$^{+}$ & 26 & ZEPLIN \cite{zepmax}, XENON \cite{xenon}, XMASS \cite{xmass2}, DRIFT \cite{drift}\\
  $^{131}$Xe & 54 & 3/2$^{+}$ & 21 & ZEPLIN \cite{zepmax}, XENON \cite{xenon}, XMASS \cite{xmass2}, DRIFT \cite{drift}\\
  $^{133}$Cs & 55 & 7/2$^{+}$ & 100 & KIMS \cite{kims}\\
  $^{183}$W & 74 & 1/2$^{-}$ & 14 & CRESST-II \cite{papcresstII}\\
  $^{209}$Bi & 83 & 9/2$^{-}$ & 100 & ROSEBUD \cite{rosebudII}\\
\end{tabular}
\end{ruledtabular}
\end{table*}

The CDMS results derive from the use of the nonzero momentum
transfer analysis of Ref. \cite{Savage} (the CDMS/$Si$ result, not
shown, constitutes a near-vertical band at $|a_{n}| \sim 1.5$,
overlapping to some extent the results of EDELWEISS and DAMA/Xe-2).
Note that the zero momentum transfer approximation does not simply
set the nuclear structure form factor appearing in the differential
WIMP-nucleus scattering rate to 1: the calculation of $
\sigma_{A}^{lim}$ involves dividing the experimental upper limit on
the WIMP rate by the convolution over the detector recoil energy
range of the form factor with the average inverse WIMP velocity.
Some experiments use a form factor independent of $a_{p,n}$ as
suggested in Ref. \cite{Lewin}, such that $\sigma_{A}^{lim}$ is also
independent of $a_{p,n}$. Other experiments (such as
\cite{naiad2,damasumma}) employ form factors dependent on $a_{p,n}$
(\textit{e.g.}, those of Ref. \cite{Ressell}) and the $
\sigma_{A}^{lim}$ in Eq. (\ref{cross}) is not the same for
$\sigma_{p}^{lim(A)}$ as for $\sigma_{n}^{lim(A)}$. When
$\sigma_{p,n}^{lim(A)}$ have been published
\cite{naiad,naiad2,ZEPLINI}, it is straightforward to use them in
Eq. (\ref{elli}) to obtain zero momentum transfer exclusions.
Provided that the form factor is not changed, changing to a nonzero
momentum transfer analysis of the same data leaves the ($a_{p}$,0)
and ($0,a_{n}$) points fixed, while rotating the major axis of the
ellipse, generally towards the nearest coordinate axis because the
absolute value of the coefficient of $a_{p}a_{n}$ is lowered. In
particular, for a single sensitive nucleus the coefficient of
$a_{p}a_{n}$ is generally less than twice the geometric average of
the coefficients of $a_{p,n}^{2}$. This removes the degeneration of
the ellipses to infinite bands predicted by the zero momentum
transfer framework for single nuclei experiments.

Fig. \ref{SDexc} includes the results from CRESST-I/$Al_{2}O_{3}$
\cite{cresst}, several recently-reported fluorine-based experiments,
and heavy nuclei searches normally considered spin-INdependent, such
as EDELWEISS \cite{edel}, ZEPLIN-I \cite{ZEPLINI} and CDMS
\cite{cdms06SD}; 50 GeV/c$^{2}$ is chosen since it lies near the
maximum sensitivities of the various experiments: for larger or
smaller $M_{W}$, all results are generally less restrictive, and
vary differentially. Each experiment is identified with the full
detector exposure in achieving the limit, rather than the
normally-quoted effective exposure (spin-sensitive detector mass
$\times$ measurement time), in order to make clear the difference
between detectors with 100\% spin sensitivity material and those
with less. Also note that the results of the indirect searches
\cite{superK,baksan}, which have been recently used to set very
restrictive limits on the WIMP-proton coupling \cite{Savage,ulka},
are not included. The 3$\sigma$ C.L. observation of the DAMA/$NaI$
annulus, appearing as two shaded bands, is taken from Ref.
\cite{danai}, which uses the standard halo model and Nijmegen form
factor, spin matrix elements \cite{Ressell}. Although this report is
from only a 159 kgy exposure, the most recent DAMA/$NaI$
\cite{damasumma} result confirms the same amplitude and phase of the
annual modulation, simply refining the error bars; as a consequence,
the shell decreases in thickness without shrinking.

As evident from Fig. \ref{SDexc}, the intersection of any two search
results, one of which is predominantly $a_{p}$-sensitive and the
other $a_{n}$, yields more restrictive limits than either of the two
alone. Clearly, the spin-independent group of experiments is
efficient in reducing the allowed spin-dependent parameter space,
despite the small (7.8\%) component of spin-sensitive $^{73}Ge$
isotope in the case of CDMS and EDELWEISS (see Table I). In Fig.
\ref{SDexc}, the predominantly WIMP-neutron sensitivity of CDMS/$Ge$
is seen to reduce the range of $|a_{n}|$ allowed by NAIAD by more
than a factor 30 (with a small reduction in $|a_{p}|$),
corresponding to the cross section limits of $\sigma _{p}\leq 0.320$
pb; $\sigma _{n} \leq 0.166$ pb obtained via Eq. (\ref{cross2})
rewritten for a single nucleon.

The future thrust of direct search activity is defined by a number
of traditionally-classified spin-independent project upgrades,
including ZEPLIN-MAX/$Xe$ \cite{zepmax}, CRESST-II/$CaWO_{4}$
\cite{papcresstII}, LIBRA/$NaI$ \cite{libra}, EDELWEISS-II/$Ge$
\cite{edelII}, GENIUS/$Ge$ \cite{genius2}, superCDMS/$Ge$
\cite{supercdms}, HDMS/$^{73}Ge$ \cite{hdms}, KIMS/$CsI$
\cite{kims}, WARP/$Ar$ \cite{warp} and ELEGANT VI/\textbf{$CaF_{2}$}
\cite{libra}. New high profile projected activity includes
XENON/$Xe$ \cite{xenon}, XMASS/$Xe$ \cite{xmass2}, EUREKA
(CRESST-II+EDELWEISS-II) \cite{eureka}, COUPP/$CF_{3}I$
\cite{coupp}, CLEAN/$Ne$ \cite{clean}, DRIFT/$CS_{2}$ \cite{drift},
ArDM/$Ar$ \cite{ardm}, DEAP/$Ar$ \cite{deap} and MIMAC/$He$
\cite{mimac}. It is not our point to review in detail these efforts:
descriptions exist as indicated and elsewhere
\cite{libra,gaitskell}. Suffice it to mention, with the exception of
the light noble liquid projects, all are "heavy" in the sense of A.
Most all of the cryogenic activities envision eventual detector
masses of up to 500 kg; the noble liquid activities, 1-10 ton. As
evident, the new activity emphasis appears to have shifted from
cryogenic searches to scintillators employing noble liquids,
reflecting a shift from phonon+ionization to
ionization+scintillation discrimination techniques in identifying
and rejecting backgrounds, as well as providing directional
sensitivity. The current background levels are \textbf{\textit{$\sim
10^{-1}$}} evt/kgd; projections for the new devices range to $\sim$
$10^{-2}$ evt/kgy. Generally, the bolometers have a few-keV recoil
threshold capacity, in contrast to the $>$ 10 keV thresholds of the
noble liquid experiments; since the Na presence permits DAMA/$NaI$
to observe a signal below the Ge recoil thresholds of CDMS and
EDELWEISS, the strong reduction of the spin-independent parameter
space still compatible with the DAMA/$NaI$ signal disappears at
masses below $\sim 20$ GeV/c$^{2}$. The recent CDMS Si-based
measurement \cite{cdms06SI} further reduces this region by more than
a factor two, but does not yet eliminate it.

\section{DISCUSSION}

The spin-dependent nuclei of the above experiments are shown in
Table I. As evident, not all of the above experiments will
contribute to further constraining this sector, in particular those
based on argon which lacks spin-sensitive isotopes.

\begin{figure}[h]
   \includegraphics[width=8 cm]{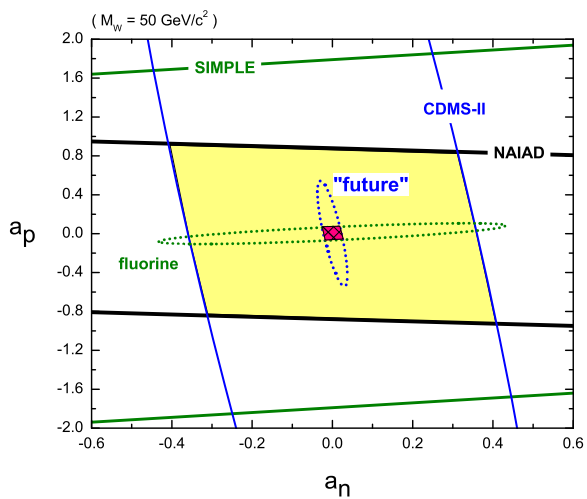}\\
   \caption{general projections (at $M_{W}$ = 50 GeV/$c^{2}$) of results
   to be expected from the current experimental direct search activity,
   in comparison with the current $a_{p}$-sensitive experimental results.
   The near-vertical ellipse denoted by "future" and suggested
   by phase A of superCDMS \cite{supercdms} indicates the general
   improvement to be achieved by the $a_{n}$-sensitive experiments discussed
   in the text. The near-horizontal ellipse indicates a similar projection
   for the fluorine-based experiments (obtained from a 200 kgd projection of the
   current SIMPLE result), with the croshatched area indicating the intersection
   of the two; the crosshatched region of Fig. \ref{SDexc} is now shown as shaded.}
\label{projSDexc}
\end{figure}

The problem with the above activity thrust for spin-dependent
investigations is shown in Fig. \ref{projSDexc}, with the projected
experimental limits of the $|a_{n}|$ sensitive experiments (ZEPLIN,
CRESST-II, EDELWEISS-II, superCDMS, HDMS, XENON, XMASS) subsumed
under the label "future" suggested by a superCDMS projection
\cite{supercdms}, and intended only to serve as an indication of the
general impact to be anticipated from the above activity (note the
change in the $a_{n}$ scale). The currently allowed area in the
parameter space of Fig. \ref{SDexc} is indicated by the shaded area,
and suggests the reduction in the allowed $a_{p}$ - $a_{n}$ space to
be achieved with the "future" thrust: generally, the limiting
ellipses of all will shrink in both parameters, but with the bounds
on $|a_{p}|$ still an order of magnitude larger than $|a_{n}|$.

\begin{figure}[h]
   \includegraphics[width=8 cm]{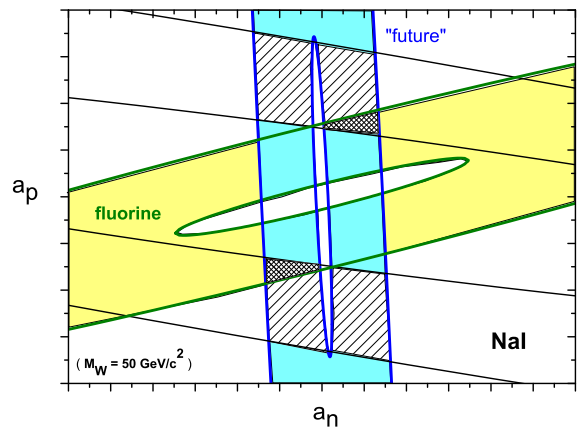}\\
   \caption{schematic of a positive result in one of the "future"
   experiments (shaded), intersecting (hatched) that of a positive $NaI$
   result. The parameter space shown as crosshatched represents the area
   allowed by the intersection of $NaI$, "future" and a similar
   positive result from one of the fluorine-based experiments. }\label{posSDexc}
\end{figure}

The point of any search experiment is however discovery, which will
be exacerbated should one or more of the spin-independent "future"
experiments obtain a positive signal. This is illustrated
schematically in Fig. \ref{posSDexc} for the case of two
"discoveries", NaI and "future". The four allowed $a_{p} - a_{n}$
regions (shaded), defined by the intersection of the hypothetical
new positive result from the "future" experiments with that of $NaI$
(corresponding to two regions of $\sigma_{p} - \sigma_{n}$), will
require at least one additional and different detector experiment of
sufficient sensitivity to further reduce the parameter space to two
allowed areas corresponding to a single pair of cross sections.

Both NAIAD \cite{naiad2} and DAMA/$NaI$ are ended. As evident,
without some additional effort, the direct search restrictions on
$a_{p}$ would remain essentially unchanged from those provided by
these measurements. The DAMA/$NaI$ experiment has been replaced by
DAMA/LIBRA \cite{libra}, an upgrade of the $NaI$ experiment to 250
kg with improved radiopurity, running since 2003. The mass increase
however provides only a factor 2.5 decrease in the exposure
necessary to confirm the original DAMA/$NaI$ signal, with further
improvement in Fig. \ref{projSDexc} scaling as
$\sqrt[4]{\textrm{exposure}}$; R\&D is in progress for a mass
upgrade to 1 ton. A second $NaI$ experiment, ANAIS, reports an
exposure of $5.7$ kgy with a $10.7$ kg prototype \cite{anais}, and
will be eventually upgraded to 100 kg. It is however only projected
to repeat the DAMA/$NaI$ measurements for confirmation of the annual
modulation.

All of these experiments rely on pulse-shape analyses for
discrimination of backgrounds. All appear to require, relative to
the leading $a_{n}$-sensitive experiments, exceedingly large
exposures, despite active masses significantly larger than the
bolometer experiments of the a$_{n}$ sector. To further limit
$a_{p}$ via direct observation, one or more experiments with
improved WIMP-proton sensitivity is required.

\subsection{Other $a_{p}$-sensitive measurements}

The KIMS experiment, based on $CsI$, has recently reported
competitive spin-independent limits with a 247 kgd exposure. Since
its sensitivity is similar to $NaI$, we show in Fig. \ref{SDexc} the
corresponding spin-dependent constraints as recalculated from the
raw data in Ref. \cite{kims}, after first reproducing the reported
spin-independent exclusion. Lacking any calculated $<S_{p,n}>$, the
result is obtained using an odd group approximation (OGA), and the
spin-dependent form factor of Ref. \cite{Lewin}. The small
gyromagnetic ratio of $^{133}$Cs yields $<S_{p}>$ $\sim$ -0.2, near
that of $^{23}Na$ (recall that the spins enter quadratically),
although the higher J of Cs implies a factor 77\% reduction in Eq.
(\ref{cross2}). To achieve the level of NAIAD, an exposure of 25.1
kgy would be required with the current background level and
discrimination; the experiment is to be upgraded to 80 kg.

As seen in Fig. \ref{SDexc}, CRESST-I achieved \cite{cresst} a
competitive result with only a 1.51 kgd exposure of a 262 g
$Al_{2}O_{3}$ bolometer, and could achieve the level of NAIAD with a
factor 50 less exposure ($\sim 300$ kgd). It has however been
abandoned in favor of $CaWO_{4}$ (CRESST-II), which is primarily
$a_{n}$ sensitive through its naturally abundant 14.3 $\%$
$^{183}W$, 0.14\% $^{43}Ca$ and 0.038\% $^{17}O$ isotopes
\cite{cresstII}. The main spin-dependent sensitivity derives from
the almost negligible $^{43}Ca+^{17}O$, the small gyromagnetic ratio
of $^{183}$W pointing to a negligible spin-dependent OGA sensitivity
($\langle S_{n} \rangle \sim$ 0.031); the current CRESST-II result
lies near $|a_{n}| \leq$ 20, well outside Fig. \ref{SDexc}.

Several activities based on new prototype devices have been recently
reported. ROSEBUD includes an $Al_{2}O_{3}$ bolometer, but the
device is only $50$ g \cite{rosebudII} and assuming the same
sensitivity as CRESST-I would require almost 16 years exposure to
achieve the current NAIAD limits. A mass increase to 1 kg would
enable limits on $a_{p}$ similar to, and more restrictive than, the
current NAIAD result with a relatively short time exposure of $\sim$
0.8 y. Like CRESST-II \cite{cresstII}, ROSEBUD however pursues
scintillating bolometers to further reject backgrounds, which if
successful could yield restrictions on $a_{p}$ equivalent to those
of NAIAD with as little as a 2 kgd exposure.

ROSEBUD also pursues measurements in BGO (= $Bi_{4}Ge_{3}O_{12}$).
As seen in Table 1, although $Ge$ and $O$ are both
neutron-sensitive, $Bi$ is proton-sensitive. As with $^{183}W$
however, a small $^{209}Bi$ gyromagnetic ratio yields an OGA
estimate of $<S_{p}>$ $\sim$ -0.085. Successful scintillation
discrimination in this case could also yield results equivalent to
those of NAIAD (although a 200 kgd exposure would be required).

The Kamioka/$CaF_{2}$ scintillator experiment of Fig. \ref{SDexc}
reports a new, very competitive limit with a total 14 kgd exposure
of a 310 g device \cite{CaF}, realized via careful attention to
component intrinsic radioactive backgrounds. It surpasses both
bolometer-based Kamioka/$NaF$ \cite{naf} and Kamioka/$LiF$
\cite{lif}. The background rate is however still roughly a factor 10
higher than those of the $NaI$ experiments, and may limit the future
performance of the detector. The recently ended ELEGANT VI is being
replaced by CANDLES III, but both are primarily focused on
$\beta\beta$ decay and have yet to provide a WIMP exclusion.

The Kamioka/$CaF_{2}$ result is essentially equivalent to recent
results reported by the two superheated droplet detector (SDD)
experiments (SIMPLE/$C_{2}ClF_{5}$ \cite{plb2},
PICASSO/$C_{4}F_{10}$ \cite{newpicasso}), with significantly less
active mass and exposure owing to inherent SDD background
insensitivity. These have so far received little attention, most
likely because of only prototype results having so far been
reported, with an unfamiliar technique. Nevertheless, given their
current results, they offer significant room for rapid improvement
in parameter space restrictions. This is shown in Fig.
\ref{projSDexc} by "fluorine" for a 10 kgd exposure with background
level of 1 evt/kgd (the Kamioka/$CaF_{2}$ experiment would require
34.5 kgy exposure with current sensitivity to achieve the same
limit, requiring either significant detector mass increase and/or
improved background discrimination to remain competitive). Being
also comparatively inexpensive and simple in construct, large volume
SDD efforts may easily be envisioned (a 2.6 kg, 336 kgd exposure
PICASSO effort is in progress, which if successful will further
reduce the crosshatched area of Fig. \ref{projSDexc}).

The simplicity argument is similarly true for COUPP \cite{coupp},
which is based on a 2 kg $CF_{3}I$ bubble chamber with the
background insensitivity of the SDDs. In this case, the 10 kgd
exposure could be achieved more quickly since the $CF_{3}I$-loading
of a SDD is only 1\% in volume \cite{tomo}. The COUPP technique
however requires a significant extension of the metastability
lifetime of the refrigerant beyond previous bubble chamber
technology. This has apparently been addressed with some success
\cite{coupp}, but a first result is still lacking.

In either case, given sufficient exposure, the fluorine experiments
combined with current CDMS results have the ability to severely
constrain the currently allowed parameter space of Fig. \ref{SDexc}.

\subsection{Spin Sensitivities}

The above complementarity of various experiments of differing
orientation in the parameter space is strongly governed by the spin
matrix elements of the involved nuclei. Unfortunately, many of the
new projected experiments lack specific spin matrix element
calculations; in their absence, several of the results are obtained
from an odd group approximation (OGA). This approximation strictly
allows only a WIMP-proton or WIMP-neutron sensitivity, by assuming
the even group to be an inert spectator so that the WIMP interacts
with only the odd group of detector nucleons. This is reflected in
the traditional spin-dependent exclusion plots, in which for
$<S_{p}>$= 0, only $a_{n}$ is constrained.

In the OGA, an experiment using only odd Z isotopes cannot constrain
the WIMP-neutron coupling. Nuclear structure calculations however
show that the even group of nucleons has a non-negligible (though
subdominant) spin. An example is $^{39}$K, which surprisingly
possesses $<S_{n}>$ = 0.05 \cite{ERTO} in spite of having a magic
number of neutrons (closed neutron shell). This can be understood
because the gyromagnetic ratio of the nucleus is low compared to
that of the nucleon, indicating a dominant contribution from the
orbital angular momentum of the proton structure; when Z $\approx$ N
and large, if the protons have a high angular momentum, so also will
the neutrons in general.

\begin{figure}[h]
   \includegraphics[width=8 cm]{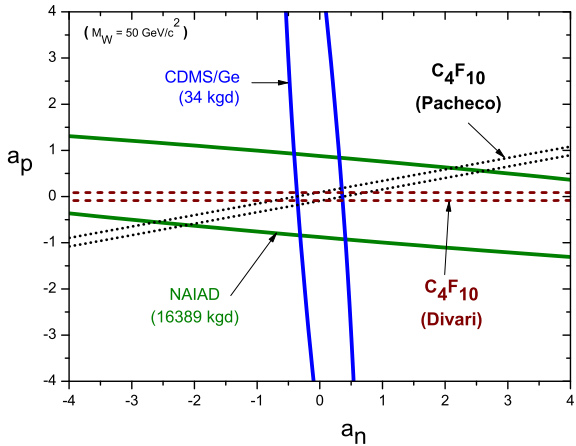}\\
   \caption{exclusion comparison for $C_{4}F_{10}$ for each of
   two sets of fluorine ($^{19}F$) spin matrix elements \cite{Pacheco,Divari},
   for a zero event 30 kgd exposure.}\label{posSDex}
\end{figure}

For A $<$ 50, the OGA has been extended by including additional
information regarding the $\beta$ decay $ft$ values and measured
magnetic moments of mirror pairs for nuclear systems \cite{Engel2},
which provides nonzero estimates of the spin matrix element for the
odd group, with seemingly small variations in the odd group spin
matrix element.

Generally, the OGA yields $<S_{p,n}>$ significantly different from
the model calculations. The refined $<S_{p,n}>$ of nuclear structure
calculations are however not measured, but obtained from various
nuclear models which reproduce known nuclear data, so that different
sets of results may exist for the same nuclide. In some cases
($^{23}$Na, $^{35}$Cl), there is even a sign reversal. Some
indication of the impact of the model difference on the contour
orientation is seen in Fig. \ref{posSDex}, for an otherwise
identical 30 kgd projection with $C_{4}F_{10}$ assuming full
discrimination.

For heavy nuclei, and/or heavy WIMPs, the zero momentum transfer
approximation breaks down and the finite momentum transfer must be
taken into account, as discussed extensively in Ref.
\cite{newnonzero}. In general this involves consideration of the
nuclear form factor ($F$) in the interaction scattering rate

\begin{equation}
        \frac{dN}{dE_{r}} \sim
        \mu^{-2}M_{W}^{-1} \sigma^{(SD)}_{A} F^{2}(q)\int^{v_{max}}_{v_{min}}
        \frac{f(v)}{v}dv   ,
\end{equation}

\noindent where $f(v)$ is related to the velocity distribution of
halo WIMPS, $v_{min}$ is the minimum incident WIMP speed required to
cause a recoil of energy $E_{r}$, $v_{max}$ is the maximum incident
WIMP speed, and $F^{2}(q) = \frac{S^{A}(q)}{S^{A}(0)}$ with the
$S^{A}$ related to the $a_{p},a_{n}$ by $S^{A}(q) =
(a_{p}+a_{n})^{2}S_{00}(q) + (a_{p}-a_{n})^{2}S_{11}(q) +
(a_{p}+a_{n})(a_{p}-a_{n})S_{01}(q)$. Calculations of the structure
functions $S_{jk}$ so far have included only $^{19}F$, $^{23}Na$,
$^{27}Al$, $^{29}Si$, $^{73}Ge$, $^{127}I$, and $^{129,131}Xe$, and
the results for the same isotope differ significantly among
calculations, depending on the nuclear potential employed
\cite{newnonzero}.

\section{CONCLUSIONS}

The future search for "spin-independent" WIMP dark matter is
particularly well-motivated and directed towards improvements of
several orders of magnitude in probing the phase space; due to the
spin sensitivity of several of the new detector isotopes, it will
also provide significant impact in the $a_{n}$ sector of the
"spin-dependent" phase space.

In contrast, the direct search in the $a_{p}$ sector is somewhat
neglected. This seems particularly strange given that the
spin-sensitivity of fluorine is well-known and that several
fluorine-based prototype experiments ($LiF, NaF, CaF_{2}$) have been
reported over the years. At present, new experiments based on
$Al_{2}O_{3}$ and fluorine are seen as possibly capable of providing
restrictions on $a_{p}$ surpassing those from $NaI$ and
complementary to those to be obtained on $a_{n}$. Of these, the SDD
and bubble chamber experiments appear to offer the greatest
possibility of achieving significantly improved restrictions with
least exposure, given their intrinsic insensitivity to most common
backgrounds; being also relatively simple in construct and less
expensive by at least an order of magnitude, large volume efforts
are readily possible. None of these experiments however seem
receiving of attention comparable to those of the $a_{n}$ activity,
which will prove problematic should any of the latter in fact
observe a positive signal in the near future.

The projected impact of several of the new, possibly interesting
spin-dependent projects, such as $CaWO_{4}$, $BGO$ and $CsI$, suffer
from the availability of only OGA estimates of their spin values,
which constrains \textit{a priori} their orientation in the
parameter space, and could profit from more detailed nuclear
structure calculations.

\begin{acknowledgments}
F. Giuliani is supported by grant SFRH/BPD/13995/2003 of the
Foundation for Science and Technology (FCT) of Portugal. This work
was supported in part by POCI grants FIS/57834/2004 and
FIS/56369/2004 of the National Science \& Technology Foundation of
Portugal, co-financed by FEDER.
\end{acknowledgments}


\end{document}